\documentclass[aps,prl,twocolumn,superscriptaddress,showpacs,floatfix]{revtex4-1}
\usepackage{color}
\usepackage{dcolumn}   
\usepackage{bm}        
\usepackage{amssymb}   
\usepackage{amsmath}
\usepackage{gensymb}   
\usepackage{bbm}
\usepackage{hyperref}
\usepackage{graphicx}
\usepackage{array}
\usepackage{braket}
\usepackage{multirow}
\usepackage{natbib}
\usepackage{overpic}

\newcommand{\bea}{\begin{eqnarray}}
\newcommand{\eea}{\end{eqnarray}}

\newcommand{\be}{\begin{equation}}
\newcommand{\ee}{\end{equation}}
\newcommand{\bk}{{{\bf{k}}}}
\newcommand{\bK}{{{\bf{K}}}}

\newcommand{\bQ}{{{\bf{Q}}}}

\newcommand{\beal}{\begin{align}}
\newcommand{\eeal}{\end{align}}
\newcommand{\ra}{\rangle}
\newcommand{\la}{\langle}

\newcommand{\upa}{\uparrow}
\newcommand{\dna}{\downarrow}

\newcommand{\dg}{{\dagger}}
\newcommand{\pdg}{{\phantom\dagger}}

\begin{document}
\title{Octupolar order in $d$-orbital Mott insulators}
\author{A. Paramekanti}
\email{arunp@physics.utoronto.ca}
\affiliation{Department of Physics, University of Toronto, 60 St. George Street, Toronto, ON, M5S 1A7 Canada}
\affiliation{International Centre for Theoretical Sciences, Tata Institute of Fundamental Research, Bengaluru 560089, India}
\author{D. D. Maharaj}
\affiliation{Department of Physics and Astronomy, McMaster University, Hamilton, ON L8S 4M1 Canada}
\author{B. D. Gaulin}
\affiliation{Department of Physics and Astronomy, McMaster University, Hamilton, ON L8S 4M1 Canada}
\affiliation{Brockhouse Institute for Materials Research, McMaster University, Hamilton, ON L8S 4M1 Canada}
\affiliation{Canadian Institute for Advanced Research, 661 University Ave., Toronto, ON M5G 1M1 Canada}
\date{\today}
\begin{abstract}
Motivated by experimental and theoretical interest in realizing multipolar orders in $d$-orbital materials,
we discuss the quantum magnetism of $J\!=\!2$ ions which can be realized in spin-orbit coupled oxides with $5d^2$
transition metal ions.
Based on the crystal field environment, we argue for a splitting of the $J\!=\!2$ multiplet, leading to a low lying non-Kramers doublet which hosts 
quadrupolar and octupolar moments. We discuss a microscopic
mechanism whereby the combined perturbative effects of orbital repulsion and antiferromagnetic Heisenberg spin interactions leads to ferro-octupolar 
coupling between neighboring sites, and stabilizes ferro-octupolar order for a face-centered cubic lattice. 
This same mechanism is also shown to disfavor quadrupolar ordering. 
We show that studying crystal field levels via Raman scattering in a magnetic field provides a probe of octupolar order. We
study spin dynamics in the ferro-octupolar state using a slave-boson approach,
uncovering a gapped and dispersive magnetic exciton. For sufficiently strong 
magnetic exchange, the dispersive exciton can condense, leading to conventional type-I antiferromagnetic (AFM) order which can preempt octupolar order.  
Our proposal for ferrooctupolar order, with
specific results in the context of a model Hamiltonian, provides
a comprehensive understanding of thermodynamics, $\mu$SR, X-ray diffraction, and inelastic neutron scattering measurements on a range
of cubic $5d^2$ double perovskite materials including Ba$_2$ZnOsO$_6$, Ba$_2$CaOsO$_6$, and Ba$_2$MgOsO$_6$. Our proposal for 
exciton condensation leading to type-I AFM order may be relevant to materials such as Sr$_2$MgOsO$_6$.
\end{abstract}
\maketitle

Multipolar symmetry-breaking orders have been extensively discussed in $f$-orbital based lanthanide and actinide compounds, which host ions 
where spin-orbit coupling (SOC) is a dominant energy scale  \cite{MultipolarRMP2009}. For instance, the ``hidden order" state of URu$_2$Si$_2$
has been extensively investigated as potentially arising from complex multipolar symmetry breaking \cite{HauleKotliar2009,Blumberg_Science2015,Blumberg_PRL2016}.
Another well-known example is cubic NpO$_2$ \cite{SantiniNpO2_PRL2000,NpO2TripleQ_PRL2002,Fazekas_PRB2003,NpO2NMR_PRL2006}, 
where a large body of experiments have been reconciled in terms of a primary antiferro-triakontadipolar (rank-$5$ magnetic multipolar) 
symmetry breaking which drives secondary antiferro-quadrupolar order. In certain pyrochlore magnets, all-in all-out
magnetic order has been proposed to lead to ``effective octupoles'' on tetrahedra \cite{Arima2013}. Ongoing experimental \cite{Sakai_JPSJ2011,Sato_PRB2012,Nakatsuji_PRL2014} 
and theoretical investigations \cite{Hattori2016,Freyer2018,SBLee2018,Patri2019} of PrTi$_2$Al$_{20}$ and PrV$_2$Al$_{20}$ 
have also uncovered quadrupolar and ferro-octupolar orders.

Recently, unconventional multipolar orders have also been proposed in $d$-orbital metals to occur as Pomeranchuk instabilities of spin-orbit coupled Fermi surfaces
\cite{LFu_PRL2015}. Specifically, metallic oxides
which have $d$-orbital ions with large SOC, such as LiOsO$_3$ and Cd$_2$Re$_2$O$_7$, have been proposed as potential candidates to realize this physics \cite{LFu_PRL2015}.
Experiments have indeed discovered an odd-parity nematic metal in Cd$_2$Re$_2$O$_7$ below  $T_c \!\sim\! 200$\,K via optical second-harmonic generation 
\cite{Hsieh_Science2017}. Other proposed materials for hosting multipolar orders include A$_2$OsO$_4$ (with A = K,Rb,Cs) \cite{Motome2018}.
However, to the best of our knowledge, there have been no clear $d$-orbital candidates for hosting octupolar orders.
Indeed, there appears to be no microscopic understanding of what are the key ingredients to potentially stabilize such octupolar phases.

In this paper, we consider spin-orbit coupled Mott insulators having transition metal ions with total angular momentum $J\!=\!2$. We
show that such Mott insulators can exhibit competing multipolar orders,
discuss a microscopic mechanism which stabilizes a ferro-octupolar state on the face-centered cubic lattice. We show how
non-resonant Raman scattering may probe the octupolar order, and compute the dynamic spin structure factor which can be measured using neutron scattering
experiments. Our work in this paper is directly motivated by a series of recent experiments on cubic double perovskite (DP) magnets, and we discuss 
how our results apply to these materials.

\section{Background review}

Ordered DP materials, with chemical formula $A_2BB^{\prime}O_6$,  are of great interest in the context of frustrated magnetism since
the B and B$^{\prime}$ sublattices individually form networks of edge-sharing tetrahedra. When only one of these ions (say B$^\prime$) is magnetically active, it results in 
quantum magnetism on the face-centered cubic (FCC) lattice, a prototypical setting for exploring geometric frustration in Mott insulators. 
Such DP Mott insulators have been studied for various electronic fillings $d^1$-$d^5$, and we briefly review some key results below.

For $d$-orbitals in an octahedral crystal field, the $t_{2g}$ single particle levels associated with the magnetic $B^{\prime}$ ion are split by SOC, resulting in 
a four-fold degenerate, $j_{\rm eff}  \!=\! 3/2$, ground state and a 
doubly degenerate, $j_{\rm eff}  \!=\! 1/2$, excited state. The physics of such materials then depends strongly on the electronic filling, $d^1$-$d^5$,
of these $t_{2g}$ states. For the most well-studied $d^5$ electronic configuration (e.g., for Ir$^{4+}$ or Ru$^{3+}$ ions), this results in a single hole 
in a $j_{\rm eff}  \!=\! 1/2$ state \cite{BJKim_PRL2008,Plumb2014}.
Recent experimental and theoretical studies on the DP Ba$_2$CeIrO$_6$, which hosts such a $j_{\rm eff}  \!=\! 1/2$ Mott insulator on the FCC lattice,
have found evidence of magnetic ordering with a strong frustration parameter, suggesting proximity to a quantum spin liquid state \cite{Aczel2019,Revelli2019}.
Stepping down to a $d^4$ configuration, strong SOC favors a total $J_{\rm eff}=0$ singlet ground state, with a gap to all excitations \cite{Khaliullin_PRL2013},
which appears to be realized in Ba$_2$YIrO$_6$ \cite{Dey2016,Kunes2016,Yan2017,ParamekantiPRB2018}. However, if intersite
exchange competes with SOC, it can lead to magnetic ordering from exciton condensation \cite{Khaliullin_PRL2013,Svoboda_PRB2017}; 
clear experimental candidates for such an exciton condensate are yet to be found.
Further down, a $d^3$ configuration would naively be expected to form an orbital singlet state with spin $S=3/2$;
however, neutron scattering and resonant inelastic X-ray scattering experiments have found that $5d$ transition metal oxides bely this
expectation, finding magnetically ordered states with large spin gaps which clearly reveal the dominance of SOC over Hund's coupling
\cite{Taylor_PRL2017,Taylor_PRB2018,Maharaj2018}.
Skipping to $d^1$ ions, we are led to a $j_{\rm eff}=3/2$ angular momentum state. Theoretical studies of such moments on the 
FCC lattice have shown that incorporating important intersite orbital repulsion can lead to complex multipolar exchange interactions, 
stabilizing wide regimes of quadrupolar order in the phase diagram \cite{ChenBalents2010,ChenBalents2011,Svoboda_arXiv2017}
which may coexist with conventional
dipolar magnetic order, or valence bond orders \cite{RomhanyiPRL2017}. 
Indeed, recent experiments on $5d^1$ oxides, Ba$_2$NaOsO$_6$ with Os$^{7+}$ \cite{Mitrovic_NComm2017,Mitrovic_Physica2018} 
and Ba$_2$MgReO$_6$ with Re$^{6+}$ \cite{Hiroi_JPSJ2019}, have found clear evidence for multiple transitions 
associated with these distinct broken symmetries, with a higher temperature quadrupolar ordering transition followed by dipolar ordering
at a lower temperature.

Finally, we turn to the topic of our work: $d^2$ ions with an effective $J_{\rm eff}=2$ angular momentum state. In this case, previous theoretical work
has found intricate multipolar couplings as for $d^1$ filling, and broad swaths of
quadrupolar orders in the phase diagram \cite{ChenBalents2010,ChenBalents2011,Svoboda_arXiv2017}.
In this paper, in contrast to previous work, we make the case that $J_{\rm eff}=2$ quantum magnets in a cubic environment may instead support ground states 
with ferro-octupolar order. We show that this can lead to a consistent understanding of a large body of experimental data on the family of cubic DP materials
Ba$_2$$M$OsO$_6$ (with $M$ = Zn, Mg, Ca), including
specific heat, magnetic susceptibility, X-ray diffraction, powder neutron diffraction, muon spin relaxation ($\mu$SR), and inelastic neutron scattering.
We present a microscopic mechanism which leads to ferro-octupolar coupling, make predictions for how Raman scattering might uncover octupolar 
order, and compute the dynamic spin structure factor which
shows a gapped magnetic exciton. Our results point to Ba$_2$$M$OsO$_6$ DPs as rare examples of octupolar order in $d$-orbital systems.

\section{Effective Local Model}

We start from an effective $J \!=\!2$ local moment, as appropriate for $d^2$ ions arising from coupling total $L=1$ and $S=1$ for two electrons.
The most general form of the octahedral crystal field Hamiltonian for $J=2$ ions is given by \cite{Maharaj2019}
\bea
H_{\rm CEF} \!=\! - V_{\rm eff} ({\cal O}_{40} + 5  {\cal O}_{44})
\label{eq:HCEF}
\eea
Here, the Steven's operators are given by
\bea
{\cal O}_{40} &=& 35 J^4_z - (30 J (J+1)-25) J_z^2 + 3 J^2(J+1)^2 \nonumber \\
&-& 6 J (J+1), \\
{\cal O}_{44} &=& \frac{1}{2} (J_+^4  + J_-^4).
\eea
For $V_{\rm eff} > 0$, this results in a non-Kramers ground state doublet, and an excited
triplet with a gap $\Delta=120 V_{\rm eff}$. As shown in a parallel publication \cite{Maharaj2019},
working in the $| J_z = m\rangle$ basis, leads to ground state wavefunctions
\bea
|\psi_{g,\uparrow}\rangle = |0\rangle;~~~
|\psi_{g,\downarrow}\rangle = \frac{1}{\sqrt{2}} (|2\rangle + | -2 \rangle)
\eea
and excited state wavefunctions
\bea
|\psi_{e,\pm}\rangle = |\pm 1\rangle;~~~|\psi_{e,0}\rangle =  \frac{1}{\sqrt{2}} (|2\rangle - | -2 \rangle).
\eea
The ground state manifold has vanishing matrix elements for the dipole operators $(J^z, J^\pm)$, precluding any dipolar order stemming from the low energy 
doublet manifold. However, $\vec J$ can induce transitions between the ground doublet and the excited triplet, which will lead to a spin-gap $\Delta$ in the excitation spectrum.
As discussed below, incorporating inter-site AF exchange would convert this local mode into a dispersing gapped `magnetic exciton'.


We have previously shown (see Supplemental Material of Ref.\cite{Maharaj2019}) that this simple model can reasonably account for the measured entropy
and magnetic susceptibility in the $5d^2$ double perovskite Mott insulators Ba$_2$$M$OsO$_6$ (with M=Zn, Mg, Ca).
Defining pseudospin-$1/2$ operators $\vec\tau$ within the ground state doublet, we find that the $e_g$ quadrupolar operators
$(J_x^2-J_y^2) \equiv 2\sqrt{3} \tau_x$, $(3 J_z^2 - J^2) \equiv - 6 \tau_z$, while the octupolar operator 
$\overline{J_x J_y J_z} \equiv - \sqrt{3}\tau_y$ (where overline denotes symmetrization).
Thus, the ground doublet can lead to time-reversal invariant quadrupolar symmetry breaking from ordering in the $(\tau_x,\tau_z)$ plane, which would also
cause non-cubic distortions due to accompanying orbital order. Alternatively, octupolar ordering with $\la \tau_y \ra \neq 0$ will lead to spontaneously broken 
time-reversal symmetry without non-cubic distortions. Finally, if the gapped magnetic exciton is sufficiently dispersive, with a bandwidth larger than the
spin gap, it can Bose condense and lead to dipolar magnetic order.

\section{Origin of ferrooctupolar coupling}

We next consider projecting microscopic intersite interactions into the low energy doublet sector described by the pseudospin-$1/2$ operators $\vec \tau$.
We have two types of interactions to consider here: type-(I) couplings have nonzero weight in the doublet sector can be directly projected into this subspace,
while type-(II) operators which mix the doublet and triplet sectors will contribute within perturbation theory. 

Examples of type-(I) interactions may be illustrated by considering a pair of neighboring sites in the $xy$-plane which will have interactions
between the $e_g$ quadrupolar charge densities $(J_x^2-J_y^2)$ or $(3 J_z^2-J^2)$ at the two sites. These
interactions may be directly projected into the doublet sector as
\be
H_{{\rm eff},xy}^{(1)} = \sum_{\la ij\ra_{xy}} (-\gamma_0  \tau_{ix} \tau_{jx} + \gamma_1 \tau_{iz} \tau_{jz}),
\label{eq:direct}
\ee
with $\gamma_0, \gamma_1 \!>\! 0$. (The effective Hamiltonian for nearest neighbors in other planes 
 can be obtained using symmetry transformations.)

Examples of type-(II) interactions for a pair of neighboring spins in the $xy$-plane include the conventional AFM exchange 
$\gamma_m \vec J_i \cdot \vec J_j$ with $\gamma_m \!>\! 0$,
where $\vec J$ denotes the $J\!=\!2$ spin. In addition, they include
$t_{2g}$ quadrupolar interactions of the form $\gamma_2 \rho_{i,xy} \rho_{j,xy}$, where $\rho_{i,xy} = (J_{i x} J_{i y} + J_{i y} J_{i x})/2$ and
$\gamma_2  \! > \! 0$. In this case, neither $\vec J_i$ nor $\rho_{i,xy}$ have matrix elements in the low energy doublet space $|L\ra$, but they 
instead mix $|L\ra$ into the high energy triplet subspace $|H\ra$, with an energy cost $2\Delta$ since both sites $(i,j)$ get excited into the triplet sector. 
We find that the effective Hamiltonian for such neighboring spins in the $xy$-plane is given, in second order perturbation theory, by 
\bea
H_{{\rm eff}, xy}^{(2)} = - \frac{1}{2 \Delta} \sum_{\la ij \ra_{xy}} (\gamma_m \vec J_i \cdot \vec J_j + \gamma_2 \rho_{i,xy} \rho_{j,xy})^2
\eea
Projecting these operators to the doublet sector, we find
\bea
H_{{\rm eff}, xy}^{(2)} &=& - \frac{1}{2 \Delta} \sum_{\la ij \ra_{xy}} \left[ 12 \gamma_m \gamma_2 \tau_{i y} \tau_{j y} +
(6 \gamma^2_m + \frac{9}{4} \gamma^2_2) \tau_{i z} \tau_{j z}  \right. \notag\\
&+& \left. (6 \gamma^2_m -12 \gamma_m \gamma_2) \tau_{i x} \tau_{j x} \right].
\label{eq:oct}
\eea
This equation is one of the key results of our paper. The first term shows that the second order perturbation theory produces a ferro-octupolar 
coupling with strength $6 \gamma_m \gamma_2/\Delta$ from the cross-coupling of $\gamma_2$ and $\gamma_m$. 
Furthermore, assuming a hierarchy $\gamma_m \ll \gamma_{2}$, we see that the net quadrupolar 
interaction, after including the terms in Eq.~\ref{eq:direct}, involves direct and perturbative contributions which come with opposite signs,
\bea
\!\!\!\!\!\! H^{\rm Quad}_{xy} \!=\!\!\!\! \sum_{\la ij\ra_{xy}}\!\! 
\left[(-\gamma_0 \!+\! 6 \frac{\gamma_m \gamma_2}{\Delta}) \tau_{ix} \tau_{j x} \!+\!  (\gamma_1 \!-\! \frac{9}{8} \frac{\gamma^2_2}{\Delta}) \tau_{iz} \tau_{jz}\right]
\label{eq:totquad}
\eea
This partial cancellation of quadrupolar couplings may cause suppression of quadrupolar order, allowing for the ferro-octupolar coupling to dominate. 
We thus identify the key microscopic mechanism underlying ferrooctupolar ordering.
A complete theory starting from an electronic hopping model with interactions, along the lines of calculations presented in Refs.\cite{ChenBalents2010,ChenBalents2011,Svoboda_arXiv2017,Hotta_PRB2017,Motome2018}, is left for future work.

\section{Raman scattering as a probe of octupolar order}

We next turn to magnetic Raman scattering \cite{RamanRMP2007} in such systems given its usefulness as a 
probe of crystal field levels and quadrupolar order in heavy fermion compounds \cite{Blumberg2019}. We show that Raman scattering
in a nonzero magnetic field leads to a new mode in the presence of octupolar order.

Our work builds on a recent study by Patri and collaborators \cite{Patri2019} which
revealed a novel magneto-elastic coupling between the strain and
the octupolar order induced by a magnetic field, which leads to linear-in-field magnetostriction as a hallmark of 
octupolar order. They argued that octupolar ordering may also lead to a softening of certain 
phonon modes in the presence of a magnetic field, which may be detectable by Raman scattering. 
Here, by contrast, we focus on the impact of octupolar ordering on the crystal field levels themselves.

The non-resonant Raman scattering intensity is given by the expression \cite{RamanRMP2007}
\bea
{\cal I}_{\mu\nu}(\omega) = \sum_{i,f} {\cal P}_i ~ |\la f | R_{\mu\nu} |i\ra|^2~ \delta(E_f-E_i-\hbar\omega)
\eea
where $i,f$ refer to initial and final states, with corresponding energies $E_i, E_f$, and ${\cal P}_i$ is the (thermal) occupation probability of the initial state. The
Raman operator $R_{\mu\nu}$ depends on the polarizations $\hat{\varepsilon}$ of the incoming and outgoing photons which differ in frequency by $\hbar\omega$. 
Here, we will focus on the specific Raman operator  corresponding to
$\hat{\varepsilon}_{\rm in} \parallel \hat{x}$ and $\hat{\varepsilon}_{\rm out} \parallel \hat{y}$, which, on symmetry grounds, is
given by $R_{xy} = (J_x J_y + J_y J_x)/2$.

In order to explore the crystal field levels of the $J=2$ ion in the presence of octupolar order and a nonzero magnetic field $B \hat{z}$,
we modify the crystal field Hamiltonian in Eq.~\ref{eq:HCEF} as
\bea
H_{\rm loc} = H_{\rm CEF} -{\mathcal B}_{\rm oct} \sum_i \tau_{i y} - B \sum_i  J_{i,z}
\eea
where ${\cal B}_{\rm oct}$ is the ferro-octupolar Weiss field in the symmetry broken phase.
This Weiss field does not impact the excited triplet wavefunctions which cost energy $\Delta$, 
but splits the non-Kramers doublet to form eigenstates
\bea
|\phi_{\pm}\ra = \frac{1}{\sqrt{2}} \left(|\psi_{g,\upa}\ra \pm i |\psi_{g,\dna}\ra\right)
\eea
which have their energies shifted respectively by $\mp {\mathcal B}_{\rm oct}$. We focus here on the zero temperature behavior of
the Raman spectrum in this local limit.

\begin{figure}[tb]
\includegraphics[width=0.5\textwidth]{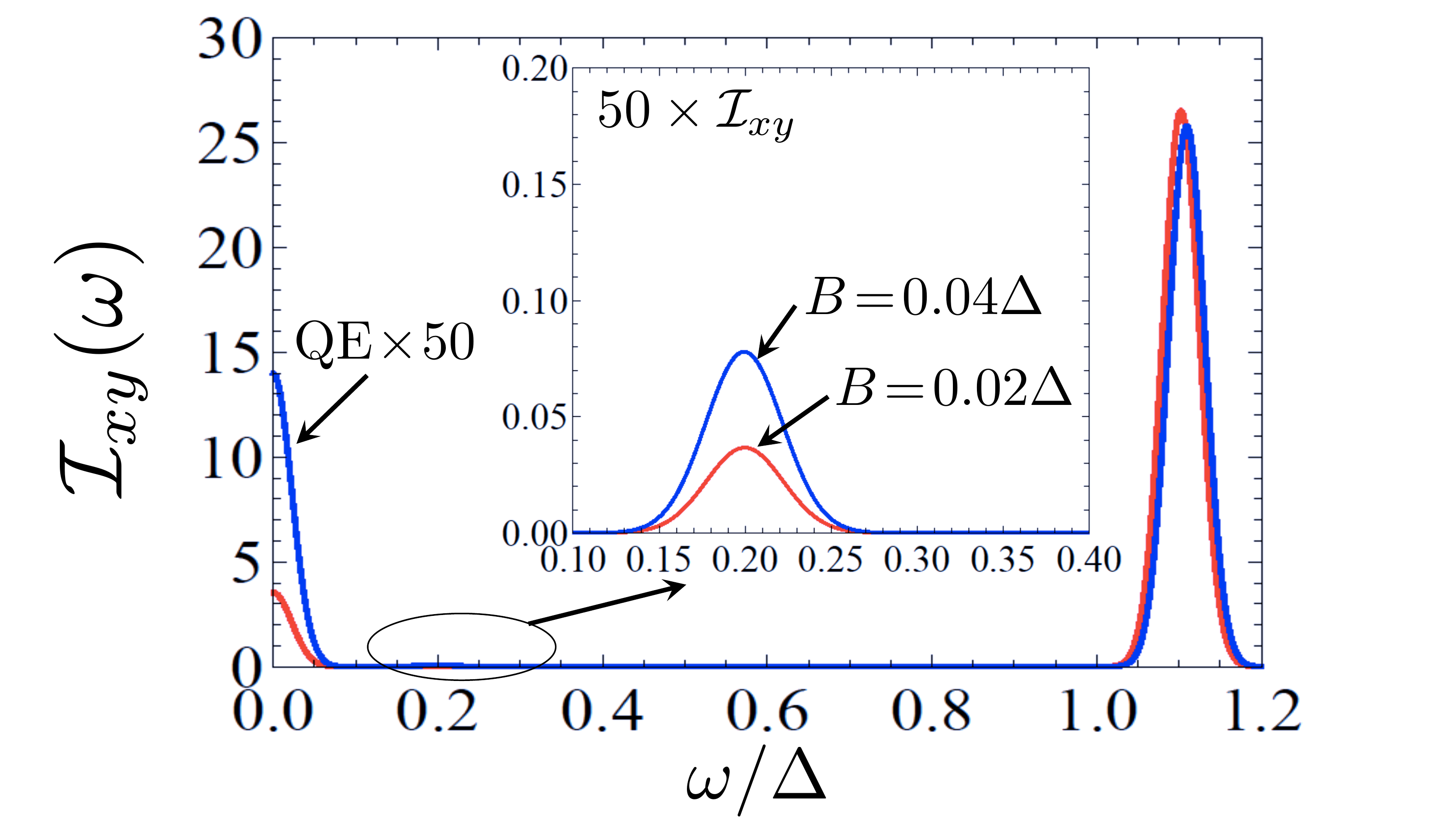}
\caption{Main panel: Raman intensity ${\cal I}_{xy} (\omega)$ as a function of frequency $\omega$ in units of the doublet-triplet
gap $\Delta$, in the presence of an octupolar Weiss field ${\cal B}_{\rm oct}=0.1\Delta$, for magnetic field $B\hat{z}$ with
$B=0.02\Delta$ (red) and $B=0.04\Delta$ (blue). The dominant peak is at $\omega=\Delta + {\cal B}_{\rm oct}$, while
QE refers to the $B$-induced quasielastic part in the presence of octupolar order,
which we have scaled up by a factor of $50$. Inset: Rescaled ${\cal I}_{xy}$ zoomed in at small low nonzero frequency,
showing an aditional mode emerging for $B \neq 0$ at a frequency $\omega \! \sim \! 2 {\cal B}_{\rm oct}$, corresponding
to the transition between the two doublet states split by octupolar order.}
\label{fig:raman}
\end{figure}

For $B\!=\! 0$, 
it is easy to show that the Raman operator $R_{xy}$ has no matrix elements in the low energy
sector $\{|\phi_{\pm}\ra\}$. Instead, at zero temperature, it induces transitions 
between the octupolar ground state $|\phi_+\ra$ and the excited crystal field state $|\psi_{e,0}\ra$ 
at energy $\Delta \!+\! {\mathcal B}_{\rm oct}$. The main panel of Fig.~\ref{fig:raman} depicts this mode which is
obtained by diagonalizing $H_{\rm loc}$ and computing ${\cal I}_{xy}(\omega)$. Here, we have artificially introduced
a broadening $\sim 0.02\Delta$ to mimic resolution effects.

Switching on $B \!\neq\! 0$ mixes the doublet and triplet wavefunctions at 
${\cal O}(B/\Delta)$, so that we must work with perturbed low energy doublet eigenstates 
\bea
|\phi'_{\pm}\ra = |\phi_{\pm}\ra \pm i \sqrt{2} \frac{B}{\Delta \pm {\cal B}_{\rm oct}} |\psi_{e,0}\ra
\eea
In addition to a weak renormalization of the above crystal field transition,
this leads to two new effects. First, we find that 
\bea 
\la \phi'_+ | R_{xy} | \phi'_{+} \ra = - 2 \sqrt{3} \frac{B}{\Delta + {\cal B}_{\rm oct}}
\eea
which is closely tied to
the linear-in-field magnetostriction explored by Patri and collaborators \cite{Patri2019} and should lead to a quasielastic
Raman signal with strength $|\la \phi'_+ | R_{xy} | \phi'_{+} \ra|^2 \propto B^2/(\Delta + {\cal B}_{\rm oct})^2$. Remarkably, a
striking parallel of such a quasielastic field-induced mode was discussed long ago in the context of (resonant) $B_{2g}$ Raman 
scattering to probe uniform scalar spin chirality in insulating square lattice antiferromagnets \cite{ShastryPRL1990,footnote.raman}.
The spin chirality breaks time-reversal but preserves spin rotation symmetry, being similar, in this sense, to 
octupolar order.

In addition, we uncover a Raman
mode corresponding to a $|\phi'_+ \ra \to |\phi'_{-}\ra$ transition at an energy $\approx 2 {\mathcal B}_{\rm oct}$.
For $B \ll \Delta-{\cal B}_{\rm oct}$, the intensity of this mode scales  $\propto B^2 {\cal B}^2_{\rm oct} / (\Delta^2-{\cal B}_{\rm oct})^2$.
The inset of Fig.~\ref{fig:raman} depicts this mode which is
obtained by diagonalizing $H_{\rm loc}$ and computing ${\cal I}_{xy}(\omega)$.

The quasielastic signal and the mode at $2{\cal B}_{\rm oct}$ are unambiguous signatures of octupolar order. Both
features lie well within the spin gap $\Delta$.
Temporally modulating $B$, or comparing 
the Raman intensity in a field relative to the zero field spectrum, might enable one to potentially search for these signals;
however, their weak intensities renders this a potentially challenging experiment.

\section{Magnetic excitons and dynamic spin structure factor}

In order to explore, in more detail, the spin excitation spectrum at energy scales on the order of the spin gap $\Delta$,
we use a slave boson approach \cite{Sachdev_PRB1990,GangChen_J0_2018,Das_NiRh2O4_2019}. 
The ensuing results can then be compared with existing inelastic neutron scattering results on Ba$_2$$M$OsO$_6$ \cite{Maharaj2019}.
We define the ground and excited states of the low energy doublet via
\bea
|\psi_{g,\sigma} \ra &=& b^\dagger_{\sigma} |{\rm vac}\ra \\
|\psi_{e,\alpha} \ra &=& d^\dagger_{\alpha} |{\rm vac}\ra,
\eea
where $\sigma=\upa,\dna$, $\alpha=0,\pm$, and
$|{\rm vac}\ra$ denotes the boson vacuum. This requires a local constraint 
\bea
\sum_{\sigma=\pm} b^\dg_{\sigma} b^\pdg_{\sigma} + \sum_{\alpha=0,\pm} d^\dg_\alpha d^\pdg_\alpha = 1.
\eea
Excitations out of the low energy space contain at least one $b$-boson; we thus get
\bea
J^{+} &=& \sqrt{6} (b_\upa^\dg d_{-}^\pdg + d_+^\dg b_\upa^\pdg) + \sqrt{2} (b^\dg_\dna d_{+}^\pdg + d^\dg_{-} b^\pdg_{\dna}) \\
J^{z} &=& 2 (d^\dg_{0}  b^\pdg_{\dna} + b^\dg_{\dna} d^\pdg_{0})
\eea
Going beyond the simple local Hamiltonian, we
model the dispersion of the gapped spin excitations using a nearest-neighbor Heisenberg exchange
$\gamma_m \sum_{\la ij\ra} \vec J_i \cdot \vec J_j$. We
supplement this, in the ferro-octupolar symmetry
broken phase, by a uniform octupolar Weiss field: $-{\mathcal B}_{\rm oct} \sum_i \tau_{i y}$. Here,
$\tau_y \equiv -i (b^\dg_\upa b^\pdg_\dna - b^\dg_\dna b^\pdg_\upa)$, and,
without loss of generality, we can set ${\mathcal B}_{\rm oct} > 0$. The total Hamiltonian we study is thus
\bea
H_{\rm spin} = H_{\rm CEF} + \gamma_m \sum_{\la ij\ra} \vec J_i \cdot \vec J_j -{\mathcal B}_{\rm oct} \sum_i \tau_{i y} 
\eea
The Weiss field favors a ground
state Bose condensate $b^\pdg_\upa \approx 1/\sqrt{2}$ and $b^\pdg_\dna \approx i/\sqrt{2}$,
resulting in the simplified expressions
\bea
J^{+} & \approx & \sqrt{3} (d^\dg_{+} + d^\pdg_{-})  - i (d^\pdg_{+} - d^\dg_{-}) \\
J^{z} &\approx & i \sqrt{2} (d^\dg_{0} - d^\pdg_{0})
\eea
Using these and accounting for the local doublet-triplet gap, we transform to momentum space, so 
the full Hamiltonian for describing the magnetic excitons is given by
\bea
H_{\rm exc} &=& (\Delta + {\cal B}_{\rm oct}) \sum_{\bk \alpha} d^\dg_{\bk,\alpha} d^\pdg_{\bk,\alpha} 
+ \frac{\gamma_m}{2} \sum_{\bk}  \eta_\bk J^+_\bk J^-_{-\bk} \notag \\
&-& \gamma_m \sum_{\bk} 
\eta^\pdg_\bk (d^\dg_{\bk,0} \!-\! d^\pdg_{-\bk,0}) (d^\dg_{-\bk, 0} \!-\! d^\pdg_{\bk,0})
\eea
where $J^+_\bk \equiv \sqrt{3} (d^\dg_{\bk,+} + d^\pdg_{-\bk,-}) - i (d^\pdg_{-\bk,+} - d^\dg_{\bk,-})$,
${J^-_{- \bk}}^\pdg \equiv (J^+_\bk)^\dg$, and $\eta_\bk = \sum_{\delta} {e}^{i\bk\cdot\delta}$ with $\delta$
labelling the $12$ nearest-neighbor vectors on the FCC lattice. This leads to a three-fold degenerate magnetic
exciton with energy dispersion given by
\bea
\lambda(\bk) = \sqrt{(\Delta + {\cal B}_{\rm oct}) (\Delta + {\cal B}_{\rm oct} + 4 \gamma_m \eta_\bk)}
\eea
We find that the exciton energy $\lambda(\bk)$ is largest at the $\Gamma$ point, and is softest at the typical FCC lattice type-I
AF ordering wavevector ${\bf K}$. 

\begin{figure}[tb]
\centering
\includegraphics[width=0.5\textwidth]{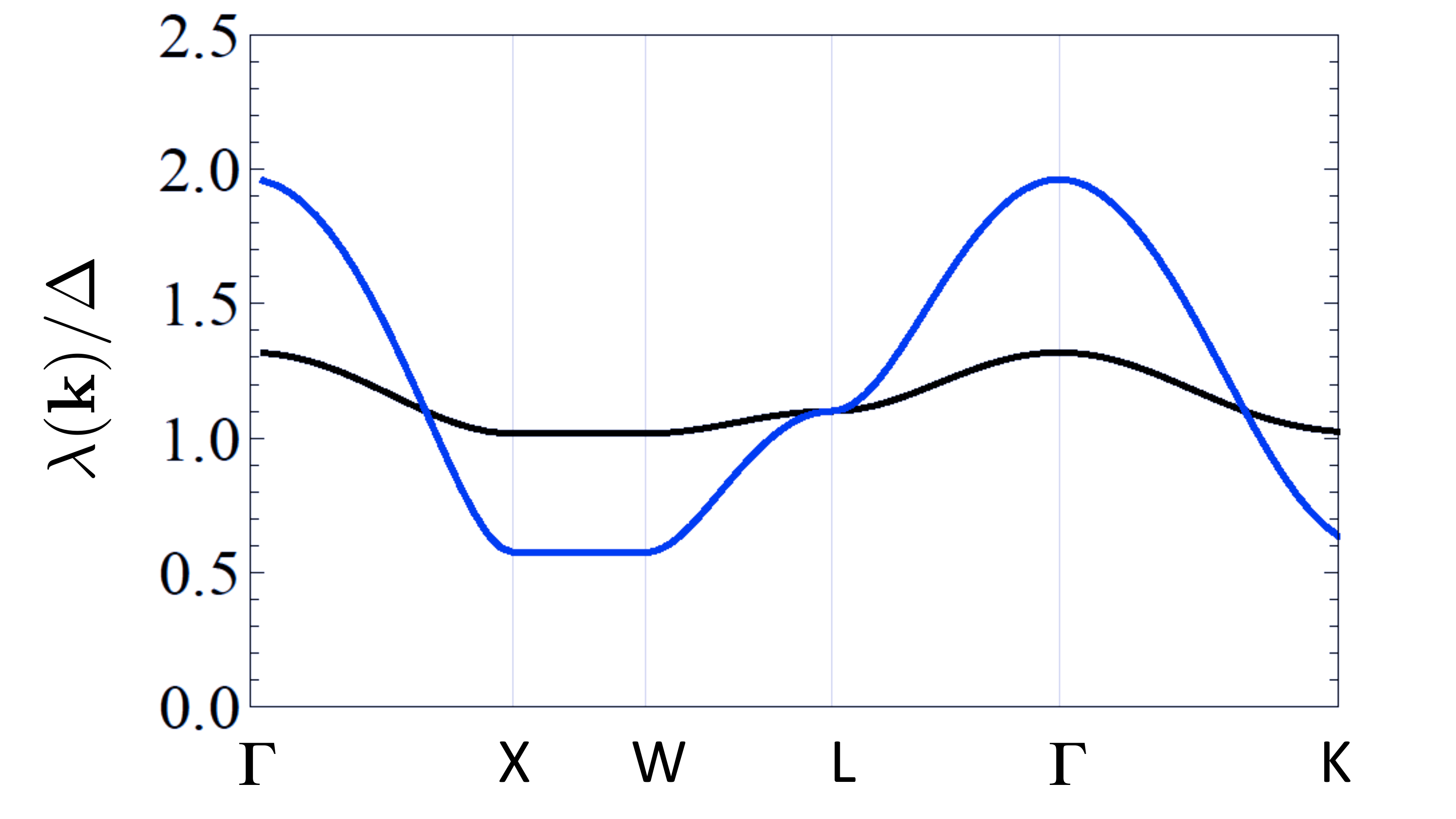}
\caption{Magnetic exciton dispersion $\lambda(\bk)$ (in units of $\Delta$) along high symmetry path in the FCC 
lattice Brillouin zone, for a choice of octupolar Weiss field ${\cal B}_{\rm oct} = 0.1\Delta$, 
and two different choices for the Heisenberg coupling: (i) $\gamma_m=0.01 \Delta$ (black) and (ii) $\gamma_m=0.05 \Delta$ (blue). 
The exciton mode clearly softens with increasing $\gamma_m$.}
\label{fig:exciton}
\end{figure}

We expect the exciton dispersion will have temperature dependence
through the temperature dependence of the octupolar order parameter, which enters via the Weiss field ${\cal B}_{\rm o}(T)$, 
softening somewhat as we heat up towards the octupolar ordering transition.
A plot of the dispersion along a high symmetry path in the FCC Brillouin zone, 
for a choice ${\cal B}_{\rm oct}/\Delta =0.1$ and $\gamma_m/\Delta=0.05$,
is shown in Fig.~\ref{fig:exciton}.
For
sufficiently large exchange coupling $\gamma_m$, the magnetic exciton can potentially condense, leading to
coexistence of dipolar and octupolar orders. The dipolar order can even preempt 
octupolar order if $\gamma_m > \Delta/16$, leading to conventional type-I AFM order.

\section{Experimental implications}

The cubic osmates Ba$_2$$M$OsO6 (with $M$ = Zn, Mg, Ca) potentially provide a realization of $J=2$ ions on the FCC lattice.
They all exhibit a single phase transition at $T^* \!\sim\!30$-$50$\,K, across which
the entropy release is only $\sim \ln(2)$ per Os, suggesting that the full $\ln(5)$ entropy is partially quenched for $T \lesssim 100$\,K \cite{Thompson_JPCM2014,Kermarrec2015,MarjerrisonPRB2016} without any phase transition.
Indeed, the structure appears to be perfectly cubic, in the $Fm\bar{3}m$ space group, at all temperatures;
both neutron diffraction  and high resolution XRD measurements find no signs of any non-cubic distortions \cite{Maharaj2019}.
This suggests that the entropy quenching above the phase transition at $T^*$ must arise from symmetry-allowed crystal field effects, 
as discussed in our theory with a non-Kramers ground state doublet.

Below the phase transition at $T^*$, neutron diffraction sees no ordered moment, even for $T \ll T^*$, instead placing tight
upper bounds on the ordered dipolar moment, $\lesssim 0.06$-$0.13 \mu_B$, depending on the material \cite{Maharaj2019}. 
At the same time, $\mu$SR measurements
have found evidence for zero field oscillations, showing spontaneous breaking of time-reversal symmetry \cite{Thompson_JPCM2014}.
Since neutron diffraction strongly hints at the absence of
dipolar magnetic order, and the cubic structure appears inconsistent with quadrupolar order, we argue that octupolar ordering within
the doublet, which preserves cubic crystal symmetry and breaks time-reversal symmetry, provides the simplest explanation for the data.
(Unlike for the $\Gamma_5$ multipoles in NpO$_2$ \cite{MultipolarRMP2009} which undergo triple-$\bQ$ ordering,
the ferro-octupolar ordering here is not symmetry constrained to induce
secondary quadrupolar order.)

\begin{figure}[tb]
\centering
\includegraphics[width=0.5\textwidth]{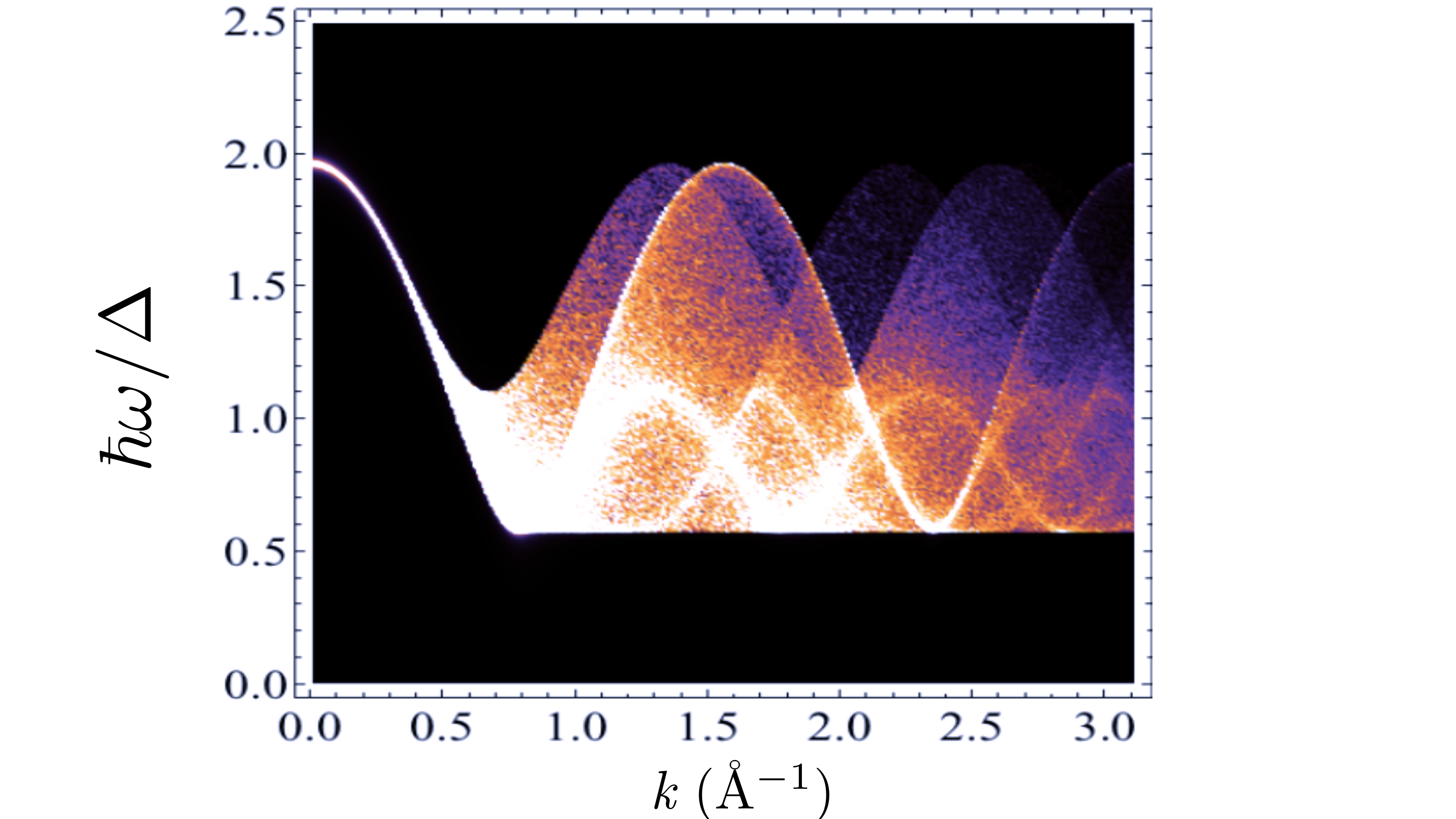}
\caption{Powder averaged dynamic spin structure factor (arbitrary units) including the Os$^{6+}$ form factor, showing the gapped magnetic exciton, 
as a function of momentum transfer $k$ (in \AA$^{-1}$, for a lattice constant $a=4$\AA) and energy $\hbar\omega$ (in units of $\Delta$). We have
set the Heisenberg exchange coupling $\gamma_m=0.05 \Delta$ and octupolar Weiss field ${\cal B}_{o}=0.1\Delta$.}
\label{fig:excitonb}
\end{figure}

Inelastic neutron scattering experiments \cite{Maharaj2019} find a spin gap $\sim 10$-$15$\,meV, 
which we interpret as arising from the doublet-triplet gap of the crystal field levels. We assume the local gap $\Delta \sim 25$\,meV, since
this yields a reasonable temperature scale below which the single-site entropy saturates to $\ln(2)$. This value of $\Delta$ is somewhat
larger than the above quoted spin gap, seen using neutrons, near the $(100)$ wavevector, but we attribute this difference
to the dispersion of the magnetic exciton.

While the measured magnetic susceptibility \cite{Thompson_JPCM2014,Kermarrec2015,MarjerrisonPRB2016} for $T \!\lesssim\! 300$\,K
in these materials hints at a Curie-Weiss temperature scale $\sim - 150$\,K, we have shown that the true $\Theta_{\rm CW}$ must be shifted
by $\sim 0.1\Delta$ due to the local spin gap, so we estimate $\Theta_{\rm CW} \!\sim\! - 120$\,K; dividing this by $z J^2$, with the FCC coordination number
$z=12$ and moment size $J=2$, we crudely estimate $\gamma_m \!\sim\! 0.25$\,meV.

Next, in order to explain $T^*$ for the Ising octupolar symmetry breaking, we must estimate the octupolar coupling constant in Eq.~\ref{eq:oct}.
We do not have any microscopic estimate for $\gamma_2$. Assuming $\gamma_2 \! \ll \! \Delta$, so that this inter-site coupling 
is weaker than the on-site CEF splitting $\Delta$, if we set $\gamma_2 \!\sim\! 5$\,meV,
we find the Ising ferro-octupolar exchange $6 \gamma_m \gamma_2/\Delta \sim 7$\,K.  Using a
classical FCC Ising model to describe the ferrooctupolar order, the known results for the Ising transition temperature \cite{Essam1963},
lead us to estimate an ordering temperature $T^* \sim 70$\,K, somewhat larger than the experimental result.
(We note that although we have explored the detailed consequences for $\gamma_2 > 0$, we are unable to rule out the possibility that $\gamma_2 < 0$ 
which would favor antiferro-octupolar order. In this case, a larger value of $|\gamma_2| \!\sim\! 50$\,meV would be necessary to explain 
the octupolar ordering temperature $T^*$; however, it is not then clear why the $\tau_x \tau_x$ coupling in Eq.~\ref{eq:totquad} would 
not cause a leading quadrupolar instability. Moreover, we do not have a microscopic explanation for such an antiferro-octupolar coupling.)

Turning to the measured exciton gap from inelastic neutron scattering, if we assume a Weiss field ${\cal B}_{\rm oct} \!\sim\! 2.5$\,meV 
(which is $\sim\!T^*/2$),
then using the above $\Delta, \gamma_m$, we find $\lambda(\bK) \!\sim\! 25$\,meV, larger than the 
measured exciton gap at $\bK$. Choosing a larger $\gamma_m \sim 1$\,meV leads to $\lambda(\bK) \!\sim\! 14$\,meV, in
better agreement with the data. These
uncertainties in $\gamma_m$ might reflect the possibility that other magnetic exchange terms could be important, 
beyond a single isotropic Heisenberg coupling.
Fig.~\ref{fig:excitonb} shows the dynamical spin structure factor 
\bea
{\cal S}(\bk,\omega) \propto \sqrt{\frac{{\cal B}_{oct}+\Delta}{{\cal B}_{oct}+\Delta + 4 \gamma_m \eta_\bk}} \delta(\hbar\omega-\lambda(\bk)),
\eea
plotted after powder averaging, and including the Os$^{6+}$ form factor. We find a high intensity gapped
band in an energy window $\sim\!(0.5 \Delta, \Delta)$, with the largest intensity concentrated at $k=\pi/a$, which corresponds
to type-I ordering wavevector $\bK=(\pi/a,0,0)$. We have assumed the Os-Os distance to be $a \sqrt{2}$, with
$a\!=\!4$\AA\, as the typical cubic lattice constant for such perovskite crystals.

We finally note that for a smaller gap $\Delta$ and stronger inter-site 
exchange, the octupolar order can coexist with dipolar order or even 
be totally preempted by Bose condensation of the magnetic exciton. The resulting
conventional type-I AFM state can have a small ordered moment if it is close to the exciton condensation transition.
We propose this scenario for Sr$_2$MgOsO$_6$ which appears to have a smaller $\Delta$ (based on its entropy) and 
a larger $\gamma_m$ (based on its Curie-Weiss temperature), and 
exhibits a type-I AFM ground state with an ordered moment $\sim\! 0.6\mu_B$, much smaller than the moment
size $\sim\! 1.88\mu_B$ inferred from high temperature susceptibility measurements \cite{Morrow_SciRep2016}. 
Weak tetragonal deformation in
Sr$_2$MgOsO$_6$ will split
the non-Kramers doublet and partially the triplet.  In this case, the magnetic exciton
condensation proposed here might still be of some relevance. However, we note that strong deformation may
partially suppress the contribution of the orbital angular momentum, and drive the system
closer to an orbitally quenched $S=1$ magnet.

\section{Summary} 

We have presented arguments in this work in favor of octupolar ordering of $J\!=\!2$ ions on the FCC lattice which
is relevant to a family of complex $5d^2$ oxides, and identified a microscopic mechanism for generating ferro-octupolar coupling.
Further theoretical and experimental work, perhaps using magnetostriction as discussed in Ref.\cite{Patri2019}, or magnetic Raman
scattering as explored here, is 
needed to provide smoking gun signatures
of the ferro-octupolar symmetry breaking. For Raman scattering, our illustrative field scale $B=0.04\Delta$ corresponds to 
$\sim 10$\,Tesla. Such experiments may require single crystals of suitable sizes. It may also be useful to
carry out more detailed microscopic calculations to compute the sign of the octupolar exchange; as noted above, the possibility
of antiferro-octupolar order is not ruled out by our work. Another interesting experimental direction would
be to apply pressure on the cubic DPs discussed here in an attempt to induce Bose condensation of the magnetic excitons. 
Our finding of a perturbative microscopic mechanism to inducing octupolar couplings via excited crystal field levels through
the combination of inter-site orbital repulsion and Heisenberg spin exchange is general
enough to be applicable to other lattice geometries. However, the specific competition between quadrupolar and octupolar
orders will depend on details of the crystal structure. Finally, this mechanism we have identified may also be of potential
importance in heavy fermion compounds, where underlying inter-site orbital repulsion and spin exchange interactions needed
to drive ferrooctupolar exchange can be induced
via coupling to conduction electrons.

\section{Acknowledgments}
This work was supported by the Natural Sciences and Engineering Research Council of Canada. AP also acknowledges
support from a Simons Foundation Targeted Grant to ICTS-TIFR.

\bibliography{octupolar}

\end{document}